%% file: async-nds.tex
\documentclass[a4paper,12pt]{article}

\usepackage[utf8]{inputenc}
\usepackage[unicode=true,hypertexnames=false]{hyperref}
\usepackage{amsmath,amssymb,amsthm,amsfonts}
\usepackage{cite}

\usepackage{tikz}
\usetikzlibrary{arrows.meta,calc,positioning}
\usetikzlibrary{external}
\makeatletter
\let\pgfmathModX=\pgfmathMod@
\usepackage{pgfplots}%
\let\pgfmathMod@=\pgfmathModX
\makeatother
\pgfplotsset{compat=1.14}

\pgfdeclarelayer{background}
\pgfdeclarelayer{foreground}
\pgfsetlayers{background,main,foreground}

\pgfplotscreateplotcyclelist{barplotcycle}{
blue!60!black,fill=blue!60!black\\
brown!60!black,fill=brown!60!black\\
orange!80!black,fill=orange!80!black\\
green!70!black,fill=green!70!black\\
red!70!black,fill=red!70!black\\
black,fill=black\\
}

\usepackage{listings}
\input{pic/tikz.tex}
\input{pic/plots.tex}

\begin{document}

\title{On Asynchronous Non-Dominated Sorting for Steady-State Multiobjective Evolutionary Algorithms
\footnote{An extended abstract of this work will appear in proceedings of Genetic and Evolutionary 
Computation Conference 2018. The paper is accompanied with the source code, the latest version of 
which is available on GitHub: \url{https://github.com/yakupov/nds2018}.}}

\author{Ilya Yakupov \and Maxim Buzdalov}

\maketitle

\begin{abstract}
In parallel and distributed environments, generational evolutionary algorithms often do not exploit the full potential of the computation system
since they have to wait until the entire population is evaluated before starting selection procedures.
Steady-state algorithms are often seen as a solution to this problem, since fitness evaluation can be done by multiple threads in an asynchronous way.
However, if the algorithm updates its state in a complicated way, the threads will eventually have to wait until this update finishes.
State update procedures that are computationally expensive are common in multiobjective evolutionary algorithms.

We have implemented an asynchronous steady-state version of the NSGA-II algorithm.
Its most expensive part, non-dominated sorting, determines the time needed to update the state.
We turned the existing incremental non-dominated sorting algorithm into an asynchronous one using several concurrency techniques:
a single entry-level lock, finer-grained locks working with non-domination levels, and a non-blocking approach using compare-and-set operations.
Our experimental results reveal the trade-off between the work-efficiency of the algorithm and the achieved amount of parallelism.
\end{abstract}

\section{Introduction}

Evolutionary algorithms, which employ evaluation of multiple candidate solutions simultaneously and independently,
are often seen as a natural choice for solving search and optimization problems 
in parallel and distributed environments. However, most evolutionary algorithms 
have two interleaving stages: the evaluation phase, where the fitness of the current population is evaluated, 
and the phase for selection and reproduction, where decisions are taken based on all the evaluated fitness values.

This design has two problems.
First, if the evaluation phase takes considerable time, and the time needed to evaluate a single individual can vary
significantly, then a large portion of available computation resources is not used while the last individuals are
waited for. Second, all the resources dedicated to fitness evaluation are idle during the second phase, 
which can also be noticeable if this phase is computationally expensive. The second issue is noticeable in algorithms
which have non-trivial state update procedures, especially if they scale asymptotically worse 
than linearly with the population size.
Most contemporary evolutionary multiobjective algorithms belong to this class,
since they contain superlinear procedures related to the maintenance of Pareto-optimal sets and 
layers~\cite{nsga-ii,nsga-iii,spea2,pesa-ii}, evaluation of indicators~\cite{ibea,hype-algorithm}
or classifying points towards reference vectors~\cite{moea-d, nsga-iii}.

Asynchronous fitness evaluation performed by steady-state evolutionary algorithms are often seen as a practical
solution to these issues. Apart from this, steady-state algorithms often have a better convergence speed than
generational ones, since each individual is sampled from a strictly better distribution than the previous one.
Studies on the steady-state variant of the NSGA-II algorithm suggest noticeable improvements over
the classic generational variant on a number of standard benchmark multiobjective 
problems~\cite{nsga-ii-steady-state}. The asynchronous implementation of the steady-state NSGA-II
has also demonstrated a better performance, in both time and diversity, for certain real-world combinatorial 
optimization problems~\cite{sync-async-moea}. Several papers also suggest that asynchronous 
steady-state algorithms have an advantage over the generational ones on problems with heterogeneous evaluation
times, either random or increasing towards the Pareto-front~\cite{nebro-durillo-master-slave-nsga-ii, 
async-moea-heterocosts, async-master-slave-moea-heterocosts}.
In several cases, within the fixed number of evaluations generational algorithms performed slightly better
in terms of the hypervolume indicator~\cite{hypervolume}, but they took considerable more time to do that
than the asynchronous algorithms~\cite{nebro-durillo-master-slave-nsga-ii}.

With bigger population sizes, however, steady-state multiobjective evolutionary algorithms tend to consume
more time in the update phase, because these procedures scale at least linearly worse than those of
generational algorithms. This problem does not manifest itself when small population sizes are used and
fitness evaluation is expensive: for instance, in~\cite{sync-async-moea} the population size ranged from
24 to 40, so the non-dominated sorting procedure from NSGA-II runs almost instantly.
However, when the population size grows, steady-state algorithms often scale worse: for instance,
the steady-state NSGA-II ran almost 10 times slower with the population size of 100
in experiments from~\cite{nsga-ii-steady-state} than the classic one.

The only part of NSGA-II which has a relatively high computation complexity is the non-dominated sorting.
This procedure is also used in the descendants of this algorithm, such as NSGA-III~\cite{nsga-iii},
and similar procedures maintain the archive of the best solutions in algorithms such as SPEA2~\cite{spea2}.
While a run of non-dominated sorting is done once for the whole population on every iteration of
a generational algorithm, in a steady-state algorithm one has to run non-dominated sorting every time
a new individual is evaluated, which is $\Theta(n)$ times slower with the population size equal to $n$.
This forced several research groups to investigate the ways to adapt non-dominated sorting algorithms
to support the incremental operations. Li~et~al.~were the first with their ENLU approach~\cite{deb-enlu-14},
see also the journal version~\cite{deb-enlu}. ENLU, or Efficient Non-domination Level Update,
handles the point addition by finding the level of the new point, comparing all points within that level
to the new one, and pushing those who are dominated to the next level. In the next level,
the points being moved are compared to all points of that level, and the new set of moving points is formed.
The worst case of one such operation is still $\Theta(n^2 k)$ for $n$ points and dimension $k$,
however, the algorithm typically runs much faster in practice. A slight improvement to one of the cases
where ENLU deteriorates was subsequently proposed in~\cite{mishra-non-dominated-level-update}.

Another line of the research was initiated by~\cite{incremental-nds-cec15}, where a faster update
procedure was proposed for the case $k=2$. Its complexity is $O(n)$ in the worst case, and it quicky
reaches $O(\log n)$ once the optimization manages to condense most of the points in the number of levels
that is at most a constant. This procedure is based on maintaining the levels as binary search trees
that can be cut or merged in $O(\log n)$. The support for $k > 2$ arrived much 
later~\cite{yakupovB-gecco17-inds}, where the algorithm is based on calling the offline non-dominated sorting
with the complexity $O(n \cdot (\log n)^{k-1})$ on two subsequent levels to push the moving points forward.
The fact that the ranks of the sorted points are known a priori made it possible to prove an improved
$O(n \cdot (\log n)^{k-2})$ worst-case bound.
The reported running times were also competitive compared to ENLU and often better.

The mentioned algorithms are not yet ready to support asynchronous multi-objective optimization
without introduction of a global lock.
However, since each update accesses levels in a sequential order, it is possible and desirable to
enrich the implementation with the possibility to introduce changes in unrelated parts by many threads 
simultaneously. This paper investigates several ways to do it. We choose the algorithm 
from~\cite{yakupovB-gecco17-inds} as the basic algorithm and developed several modifications:
apart from the obvious modification to introduce the global lock on the entire data structure,
we considered an implementation based on the compare-and-set concurrency primitives,
as well as an implementation which uses finer-grained level-based locks.
These implementations are evaluated on synthetic datasets generated by
the asynchronous NSGA-II.

\section{Preliminaries}

In this section, we briefly introduce the notation we use and the core concepts necessary to
understand this paper.

\subsection{Notation}

Without loss of generality, we consider multiobjective minimization problems.
Since in large parts of this paper we do not consider particular optimization problems or fitness functions,
we typically do not differentiate between genotypes and phenotypes, so we treat individuals as points
in the $k$-dimensional objective space.

A point $p$ is said to \emph{strictly dominate} a point $q$, denoted as $p \prec q$, if in every coordinate
$p$ is not greater than $q$, and there exists a coordinate where it is strictly smaller:
\begin{equation*}
    p \preceq q \leftrightarrow \begin{cases}
        \forall i, 1 \le i \le k, p_i \le q_i;\\
        \exists i, 1 \le i \le k, p_i < q_i.
    \end{cases}
\end{equation*}

There also exists a \emph{weak domination} relation, denoted as $p \preceq q$, which removes the second 
condition. We use the term \emph{domination} for strict domination if not said otherwise.

\begin{figure}[!t]
\centering
\scalebox{0.75}{
\begin{tikzpicture}[scale=0.28]
\TikZPictureLevel
\end{tikzpicture}
}
\caption{The working principles of incremental non-dominated sorting.
On each phase, a set of moving points is considered, which initially consists of a single
point that is inserted. The points that are dominated by the nadir of the inserted points
are selected, and the offline non-dominated sorting is performed on the union.
The points that get rank 0 remain in the current front, while others become the next
moving points.}\label{inds-demo}
\end{figure}
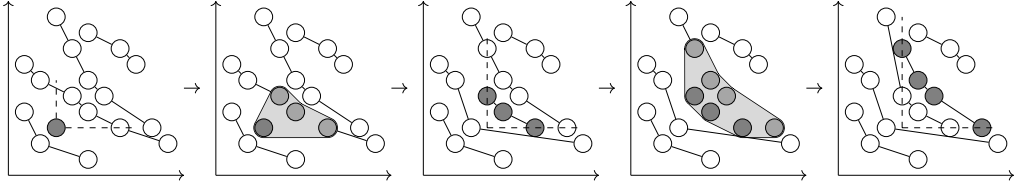

\emph{Non-dominated sorting} is a procedure that takes a set of points $P$ and assigns each point a 
\emph{rank}. The points from $P$ that are not dominated by any other points from $P$ receive rank 0.
All points that are dominated only by points of rank 0 receive rank 1. Similarly, all points
that are dominated only by points of rank $\le i$ receive rank $i + 1$.
A set of points with the same rank is called a \emph{non-domination level}, or simply a \emph{level}.
The first picture in Figure~\ref{inds-demo} shows an example of four non-domination levels of white points in 
two dimensions.

\emph{Incremental non-dominated sorting} is a procedure that updates ranks of a set of points
when a new point is inserted or deleted. There are several algorithms to perform incremental
non-dominated sorting~\cite{deb-enlu,incremental-nds-cec15,yakupovB-gecco17-inds}, of which
the one from~\cite{yakupovB-gecco17-inds} currently has the best performance among the ones
for arbitrary dimension $k$.

\emph{Crowding distance} is the quantity used for diversity management within a non-domination level
in NSGA-II~\cite{nsga-ii}. For a point $p$, the crowding distance is equal to:
\begin{equation*}
CD(p) = \sum_{i=1}^{k} \frac{p^{\text{right}}_i - p^{\text{left}}_i}{P^{\max}_i - P^{\min}_i},
\end{equation*}
where $P^{\max}_i$ is the maximum among the $i$-th coordinates in the population
(similarly $P^{\min}_i$ is the minimum), and $p^{\text{right}}$ is the point from the population
with the $i$-th coordinate just above $p_i$ (similarly $p^{\text{left}}$ is the point just below;
in other words, when the population is sorted by the $i$-th coordinate, $p^{\text{right}}$ and
$p^{\text{left}}$ are the neighbors of $p$). If at least one of the neighboring point is absent,
then $CD(p) \gets \infty$.

\subsection{Concurrency Primitives}

There exists a number of different concurrency primitives to ensure certain ordering on
operations in multithreaded environment. Maybe the simplest one is the \emph{lock}.
It is mostly used to surround a so-called \emph{critical section}: a region of the code
that is intended to be executed by a single thread only. Locks basically support two operations:
\emph{acquire} and \emph{release}. If the lock is not acquired by any thread, the first one that
calls \emph{acquire} does it and can proceed. Any subsequent thread that calls \emph{acquire}
will be suspended until the first thread releases the lock (by calling \emph{release}). When the lock
is released, one of the threads waiting for this lock will resume and acquire this lock.
A simple Java code example below shows an example of the usage of locks.

\begin{lstlisting}[language=Java]
Lock lock = new Lock();
void procedureUsingLock() {
    callSomethingThreadSafe();
    lock.acquire();
    callBySingleThreadOnly();
    lock.release();
}
\end{lstlisting}

Some programming language, including Java, introduce a more complex concept called \emph{monitors}.
However, when the special methods \texttt{wait()}, \texttt{notify()} and \texttt{notifyAll()} are not used,
they are similar to locks. One can \texttt{synchronize} on an object, which is similar to acquiring
a lock associated with that object and subsequently releasing it. A method of an object can be marked as
\texttt{synchronized}, which is essentially equivalent to synchronizing on this object for the course
of the entire method.

There is also a number of finer primitives, which are not associated with critical sections of code,
but instead guard the order in which a certain dedicated memory area is accessed or modified.
One of them is called \emph{compare-and-set}. In simple words, one can access a variable, test it
for equality to a reference value, and only if these values are equal, set the variable to another specified
value, as if this all is a single uninterrupted instruction, that is, \emph{atomically}. In a Java notation, 
it is roughly equivalent to:

\begin{lstlisting}[language=Java]
int value;
Object lock = new Object();
boolean compareAndSet(int ref, int newVal) {
    synchronized (lock) {
        if (value == ref) {
            value = newVal;
            return true;
        } else {
            return false;
        }
    }
}
\end{lstlisting}
but is typically much faster. Many modern processors do indeed provide a similar instruction,
such as the compare-and-exchange (\texttt{CMPXCHG}) instruction in the x86 family.

The compare-and-set functions can be used to implement \emph{non-blocking} algorithms,
in particular \emph{lock-free} and \emph{wait-free} algorithms, which, 
unlike the ones using locks or monitors,
do not force threads to wait one for another. Such algorithms can theoretically scale better than
the lock-based ones when the number of processors is growing.
However, a non-blocking algorithm is not guaranteed to be better, since it can perform much more
unnecessary work if not designed properly.
For instance, an efficient wait-free algorithm
for the wait-free queue
was proposed as recently as in 2011~\cite{wait-free-queues}.

\subsection{Incremental Non-dominated Sorting}

Here we briefly describe the core principles of the incremental non-dominated sorting algorithm 
from~\cite{yakupovB-gecco17-inds}. They are also illustrated in Figure~\ref{inds-demo}.

The algorithm maintains the levels in separate lists, ordered lexicographically from the first objective
to the last one. On insertion of a point $p$, first the maximum number of level $\ell$,
where a point exists that dominates $p$, is found. Then, a set of moving points $M$ is formed, initially
$M = \{p\}$, which contains a subset of the points that increase their rank.
An algorithm for offline non-dominated sorting is then run on $L_{\ell+1} \cup M$. Since both $M$ and 
$L_{\ell+1}$ are both non-dominating sets, and no point from $L_{\ell+1}$ can dominate a point from $M$,
the rank of each point will be either 0 or 1. The points of rank 0 form the new level $L_{\ell+1}$,
rank 1 forms the new $M$, and then the process continues with $\ell \gets \ell + 1$.

The existence of only two ranks, 0 or 1, may improve the performance of non-dominated sorting: for instance,
the algorithm from~\cite{jensen,buzdalov-nds-2014}, which normally runs in $O(n \cdot (\log n)^{k-1})$,
speeds up to $O(n \cdot (\log n)^{k-2})$, because the $O(n \log n)$ algorithms that form its baseline for the 
divide-and-conquer degenerate to $O(n)$ in the presence of two ranks.
Together with the fact that points from $L_{\ell+1}$ can never dominate points from $M$, this also enables 
calling directly the internal procedure of this algorithm, which assigns ranks to inferior points given that 
superior points are fully evaluated (this procedure is often called \textsc{HelperB} following the notation of
the paper which introduced the methodology~\cite{jensen}).
One more insight that further improves the performance is that we can first exclude those points
from $L_{\ell+1}$ that are not dominated by the coordinatewise minimum, or the \emph{nadir}, of points
from $M$.

\section{Introducing Concurrency}

In this section, we show two major ways for how to introduce concurrency into the incremental
non-dominated sorting. Note that there also exists a simple and inefficient way, namely,
to put all procedures that can update the levels under a single lock. In Java, one would
modify a class which representes the collection of levels by putting
the \texttt{synchronized} modifiers on all methods which query or modify the levels.
This is, however, still a valid baseline method for subsequent comparisons, along with the
single-threaded evolutionary algorithm.

\subsection{The Compare-And-Set Approach}\label{cas1}

In the approach based on compare-and-set primitives, we optimistically let the threads do their work on 
updating the levels in their local memory areas and publish the results of their computations in the
case no other thread had updated this level before. Each level is stored in its own dedicated memory area
that is updated atomically (for this purpose, in Java we use \texttt{AtomicReference} of an object that 
contains the points of the level along with the necessary metadata), so we ensure that the threads
can work with the point sets that are internally consistent (for instance, each level consists of points that
do not dominate each other).

When using this logic, however, we cannot rely anymore on the fact that the set of moving points $M$
and the level $L_{\ell+1}$ we are insertion these points into are related in such a way that no point
$p \in L_{\ell+1}$ can dominate a point $m \in M$. Indeed, since the time the current thread has formed
the set $M$ and left the previous level $L_{\ell}$ in a consistent state, another thread might have
updated the front $L_{\ell+1}$. Since every such update makes the level closer to the Pareto front by 
any sensible measure (such as the hypervolume~\cite{hypervolume}), some points can appear in $L_{\ell}$
that dominate some points in $M$.

Given this fact, we have to resort to the full-blown offline non-dominated sorting to determine the
new contents of $L_{\ell+1}$. We can use, however, the fact that the set of points to be sorted
is formed by a union of two sets, $M$ and $L_{\ell+1}$, each of which is non-dominating. It follows from
this fact that the ranks will be either 0 or 1 again, and, by induction, the next $M$ will also be
non-dominating. As a consequence, the runtime of non-dominated sorting will be
$O(n \cdot (\log n)^{k-2})$ for $k > 2$.

Once a thread has computed the new value candidates of $L_{\ell+1}$ and $M$, it performs the compare-and-set
on the atomic variable holding the actual value of $L_{\ell+1}$. If $L_{\ell+1}$ at this time
is exactly the same as before the sorting, then the update succeeds and the thread moves on with a new $M$ 
to another level ($\ell \gets \ell + 1$). Otherwise, some other thread has changed $L_{\ell+1}$ before the 
current one, so it has to perform the process again until it succeeded.

In this implementation, we use one lock to guard the relatively infrequent situations when a new level is 
added or the last level is removed.
We also have to quit using the heuristic which stops propagation of levels and creates a new one
once the set $M$ dominates the set $L_{\ell+1}$ entirely.

\subsection{A Time-Stamping Modification}\label{cas2}

To use the benefits offered by a faster merging of levels in~\cite{yakupovB-gecco17-inds},
we introduce time-stamping of levels. In this modification, each level has an associated integer number,
which is increased at the beginning of each point insertion, and also on creation of a new version of a level.
In the latter case this increased value is associated with this new version.
The timestamps originate from a single atomic integer variable global to the particular set of levels,
which can be atomically incremented when in use by multiple threads.
While performing operations associated with insertion of a certain point, we keep the time-stamp 
$\tau$ corresponding to the moment when this insertion is started.

Whenever we perform the merging of the set of moving points $M$ and the currently modified level
$L_{\ell+1}$, and the time-stamp $T(L_{\ell+1})$ is less than $\tau$, it means that this level was not
modified by any thread. In this case, the invariant that no point $p \in L_{\ell+1}$ can dominate
any point $m \in M$ holds, since $M$ consists of the points that are \emph{at least as good},
in terms of domination, as the points from $L_{\ell}$ at the time $\tau$.

Note that the above holds even if for this particular insertion there were previously several levels
for which the time-stamp was greater than $\tau$. This can indeed happen since several insertions
running in parallel could terminate earlier than the current one, or the current thread could be given
a time slot enough to overcome other threads.

\subsection{The Approach with Finer-Grained Locks}\label{lock}

We have also implemented a version which has a lock associated with each level.
When performing an update of the level $L_{\ell+1}$ by a set of moving points $M$,
the thread acquires a lock $K_{\ell+1}$ associated with the updated level.
By this it ensures that no other thread will modify the level $L_{\ell+1}$ by the time
it is done with the sorting.

Just before the lock $K_{\ell+1}$ is released, the thread acquires a lock 
$K_{\ell+2}$ associated with the next level $L_{\ell+2}$ if the new set of moving points $M$
is not empty. By doing this, the thread ensures that there will be no other thread which surpasses it.
In turn this also ensure the condition that points from $L_{\ell+2}$ cannot dominate points
from the new version of $M$.

When the locks are used in this way, threads which update the levels always follow each other
in an unchanged order in the direction of increasing of level indices.
This is a property which greatly simplifies thinking about the algorithm as well as the formal proofs.
However, this also results in many threads competing for the lock of the last level, since
a thread typically not only adds a point, but also removes the worst point, which is located in the last 
level. To partially overcome this, we do not delete points unless the number of points in all levels
exceeds $1.2 \cdot n$ for the desired population size $n$. Once this threshold is reached, the extra
$0.2 \cdot n$ worst points are removed. Since this process can require removal of a large number of levels,
a separate lock to handle this process was also introduced.

\subsection{Recomputation of the Crowding Distance}

When the algorithms for incremental non-dominated sorting are used within the NSGA-II algorithm,
they need to support querying of a point, along with its rank and crowding distance, by its ordinal
(that is, by its index in some arbitrary but predefined order). This is mostly trivial except for the crowding 
distance.

Since the crowding distance requires the knowledge of the coordinate-wise span of the level
in which the point resides, as well as the neighbors of this point in every coordinate,
the information needed to compute the crowding distance is not local. This presents an issue in the realm of 
incremental non-dominated sorting, since it typically performs small changes to the levels. In particular,
the size of the moving set $M$ is often much smaller than the size of the levels, and the subset of the level
$L_{\ell+1}$, which is dominated by the coordinate-wise minimum of $M$ is often also small.
The computationally complex non-dominated sorting is performed only on these small parts, while the remaining
part of the level $L_{\ell+1}$ is processed using a routine with the complexity $O(nk)$.
The complexity of crowding distance, if performed on the entire level, is $O(nk \log n)$,
which appears to dominate the running time of the entire algorithm.

We propose a way to reduce this part of the running time to $O(nk + \tilde{n}k \log \tilde{n})$,
where $\tilde{n}$ is the size of the small parts from above. One of the ways to do it is to maintain, in each level
and for each coordinate, a list of points contained in this level sorted in that coordinate.
After an update, for the newly inserted points the lists sorted in each coordinate are constructed,
and then these lists are merged with the lists stored in the level in $O(n)$ time each.
During these merges, the entires corresponding to the just removed points are also removed from the lists, 
and the crowding distance is recomputed for every point.

\section{Experiments}

For the experimental evaluation, we have considered the algorithms mentioned above:
\begin{itemize}
    \item INDS: the incremental non-dominated sorting algorithm from~\cite{yakupovB-gecco17-inds};
    \item Sync: the same algorithm with all public methods annotated with \texttt{synchronized};
    \item CAS1: the modification of INDS according to Section~\ref{cas1};
    \item CAS2: the modification of CAS1 according to Section~\ref{cas2};
    \item Lock: the lock-based modification of INDS according to Section~\ref{lock}.
\end{itemize}

All these algorithms were implemented
in Java (OpenJDK with the runtime version 1.8.0\_141), 
and their performance was evaluated on an 64-core machine with
four AMD Opteron\texttrademark\ Processor 6380
processors clocked at 2.5 GHz running a 64-bit GNU/Linux OS (kernel version 3.16.0).

We evaluated the algorithms on the well-known benchmark problems DTLZ1--DTLZ4 and DTLZ7~\cite{dtlz}, as well 
as ZDT1--ZDT4 and ZDT6~\cite{zdt}. For the ZDT problems, we kept $k=2$.
For all DTLZ problems, we performed experiments with $k=3$, and additionally
we ran DTLZ1 and DTLZ2 problems with $k \in \{4,6,8,10\}$. 

The datasets were synthesized for each problem as follows.
First, a random population of size 5000 was created. Then, a steady-state NSGA-II was run for
the next 1000 iterations, creating 1000 points to be inserted. The initial population, as well as the
inserted points, were recorded for the usage in benchmarking. For each DTLZ problem,
three datasets were synthesized in this way, while for the ZDT problems the number of datasets was two.

Each of the algorithms was then run on the datasets, and their running time was measured by the
Java Microbenchmark Harness framework (JMH, version 1.17.2) with four warm-up and four measurement
iterations, each at least one second long, using two independent forks of the Java Virtual Machine.
Every run consisted of initialization of the algorithm on the initial population from the dataset,
which was not counted towards the running time, and insertion of the 1000 points from the dataset,
together with the subsequent deletion of the worst point, which was measured.

For all algorithms except INDS, which is sequential by its nature, a number of threads was used to
insert the points. The number of threads was taken from $\{3, 6, 12, 24\}$.
The poits to be inserted were evenly and randomly distributed between the threads.
Figure~\ref{zdt} shows the results for all ZDT problems, which are two-dimensional.
Figure~\ref{dtlz3d} shows the results for all three-dimensional DTLZ problems.
Figure~\ref{dtlz1} is dedicated to DTLZ1 with different values of $k$,
while Figure~\ref{dtlz2} does the same for DTLZ2.

\subsection{ZDT, Two Dimensions}

The results on the ZDT problems reveal that one can not generally benefit from having
an asynchronous algorithm when the average insertion time is very small
(it is typically around $10^{-1.6} \approx 0.025$ seconds as Figure~\ref{zdt} suggests).
The Sync version shows that thread contention introduces slowdowns that are orders of magnitude
worse that the running time of the algorithm itself.

These results also show that the algorithms based on the compare-and-set mechanism scale rather well
in these conditions. There is a stable and distinct trend for the running time of CAS1 to decrease
while the number of threads increases. CAS2, due to its optimizations, is initially rather fast
and somewhat competitive with the single-threaded INDS. A minimum located somewhere
between 6 and 12 threads can be observed for CAS2.

The Lock algorithm, similar to the Sync one, degrades with the growth of the number of threads,
however, its performance is much better than of Sync, in particular, it stays competitive to INDS
when three threads are used. This behavior is generally expected from the lock-based algorithms.

\subsection{DTLZ, Three and More Dimensions}

Things, however, change in three dimensions, where the cost of a single insertion
raises towards approximately $0.07$ seconds. In these settings, the performance of Sync,
which is still much worse compared to INDS, but not to the scale observed on the ZDT problems.

The performance of CAS1, however, becomes much worse compared to even Sync.
It retains the trend towards better scaling with the number of threads, however, it is worse than
Sync even when both use 24 threads. 
The key problem with CAS1 seems that it often spends much time in sorting, which gets more time-consuming
in three dimensions, and only to find that some other thread has overwritten the target level.

CAS2, on the other hand, retains relatively efficient, however, it still does not exceed INDS in the 
performance. In this setting, it demonstrates the trend towards increasing its running time with
the number of threads. It looks like even with the improvements introduced to CAS2 the amount of work 
every thread wastes in order to recompute the level insertion once some other thread overcame it
is growing with the number of threads and is not compensated by the absence of idle time.

The biggest surprise is the Lock algorithm, which demonstrates roughly the same performance as in
two dimensions and thus overcomes INDS in the performance. Figure~\ref{dtlz3d} also shows a 
consistently better behavior with 6 threads.

The same trends are demonstrated also in higher dimensions on the DTLZ problems, which suggests
that the lock-based algorithm is an algorithm of choice in the concurrent environments, at least for this 
number of points, threads and dimensions. Its behavior regarding the number of threads seems to be quite 
robust, although there is indeed a slight trend towards increasing the running time when the number
of threads grows. A local minimum around six threads is observed for $k \le 4$, while this behavior
tends to disappear for larger values of $k$.

A possible explanation for such a good behavior of Lock can be that, after a short initial phase,
the threads start to follow each other with some short distance in the same order for long periods of time.
It is yet an open question whether it is true, and whether the picture is going to change with 
heterogeneous times of fitness evaluation.

\section{Conclusion}

We have made the first step towards efficient data structures for large-scale asynchronous 
steady-state multiobjective algorithms based on non-dominated sorting. Our experiments
suggest that an asynchronous implementation of incremental non-dominated sorting with
fine-grained level-based locking seems to be a viable choice already at population size of several
thousand points with dimensions starting from $k = 3$. We should, however, notice that
the benefits from using more threads for insertion of points are not very clear, although
the algorithm seems to tolerate our tested maximum of 24 threads pretty well.

It also looks like more advanced approaches, such as the algorithms based on the compare-and-set primitives, 
are more difficult to be made practical, at least with the chosen design of such algorithms. 
By this paper, we did \emph{not} prove that work-efficient lock-free algorithms do not exist for
incremental non-dominated sorting. We only showed that a particular design, namely, comparing-and-setting
entire levels, is probably not very efficient. Doing this on the level of single individuals, however,
does not sound promising either, since it is very likely that this will multiply the computation costs of 
non-dominated sorting itself by the overhead of compare-and-set primitives.

An approach based on locking of individual levels seems to be somewhat natural, as it ensures that threads
walk the levels one after another. However, it is yet an open question whether the access of a single level
by multiple threads, which operate at different non-intersecting parts of that level,
can be efficiently implemented for reasonable problem sizes. It can possibly be done by checking in
$O(nk)$ whether the regions dominated by two different sets of moving points, that are manipulated by
different threads, intersect in the current level: if they do not, then this level, and all subsequent
levels, can be processed by these two threads only with minor resource sharing, since the most expensive parts
will operate with non-intersecting sets of points.

Yet another possibility, which may find its use in heterogeneous computing systems (where the 
internals of an evolutionary algorithm is run with different computation resources than fitness evaluation)
is a special flavor of an asynchronous algorithm which, on the arrival of a fitness thread,
hands the next task immediately, and only then inserts the evaluated point into the data structure.
This should reduce the idle rate of fitness-related computation resources, which are typically more expensive.
However, an impact of this design on the convergence of an algorithm, as compared to
the one implemented in this paper, may be non-trivial.

As our future work, we plan to investigate the performance of the asynchronous algorithms
in more realistic settings, such as working within a real evolutionary multiobjective algorithm,
as well as with heterogeneous times of fitness evaluation. An extension of this approach to
computing more different types of diversity measures, such as the reference-point based measure of
NSGA-III, is also worth investigating.

\section{Acknowledgments}

This work was supported by Russian Science Foundation under the agreement No.~17-71-20178.

\bibliographystyle{abbrv}
\bibliography{../../../../bibliography}

\appendix

\newcommand{\includefivewide}[6]{
\begin{axis}[ybar, bar width=3pt, 
ymode=log, cycle list name=barplotcycle, xlabel={#6},
xtick={INDS, Sync, CAS1, CAS2, Lock}, 
symbolic x coords={INDS, Sync, CAS1, CAS2, Lock}]
\addplot plot[error bars/.cd, y dir=both, y explicit] table[y error=y-dev] {#1};
\addplot plot[error bars/.cd, y dir=both, y explicit] table[y error=y-dev] {#2};
\addplot plot[error bars/.cd, y dir=both, y explicit] table[y error=y-dev] {#3};
\addplot plot[error bars/.cd, y dir=both, y explicit] table[y error=y-dev] {#4};
\addplot plot[error bars/.cd, y dir=both, y explicit] table[y error=y-dev] {#5};
\end{axis}}

\begin{figure}[!t]
\begin{tikzpicture}[scale=0.8]
\includefivewide{\zdtBdCtB}{\zdtBdCtD}{\zdtBdCtG}{\zdtBdCtBC}{\zdtBdCtCE}{ZDT1}
\end{tikzpicture}
\begin{tikzpicture}[scale=0.8]
\includefivewide{\zdtCdCtB}{\zdtCdCtD}{\zdtCdCtG}{\zdtCdCtBC}{\zdtCdCtCE}{ZDT2}
\end{tikzpicture}
\begin{tikzpicture}[scale=0.8]
\includefivewide{\zdtDdCtB}{\zdtDdCtD}{\zdtDdCtG}{\zdtDdCtBC}{\zdtDdCtCE}{ZDT3}
\end{tikzpicture}
\begin{tikzpicture}[scale=0.8]
\includefivewide{\zdtEdCtB}{\zdtEdCtD}{\zdtEdCtG}{\zdtEdCtBC}{\zdtEdCtCE}{ZDT4}
\end{tikzpicture}
\begin{tikzpicture}[scale=0.8]
\includefivewide{\zdtGdCtB}{\zdtGdCtD}{\zdtGdCtG}{\zdtGdCtBC}{\zdtGdCtCE}{ZDT6}
\end{tikzpicture}
\caption{Experiments with ZDT problems. For all the problems, $k = 2$.
For every asynchronous algorithm the columns represent the number of threads.
They are, left-to-right, 3, 6, 12, 24. Times are given in microseconds.}
\label{zdt}
\end{figure}
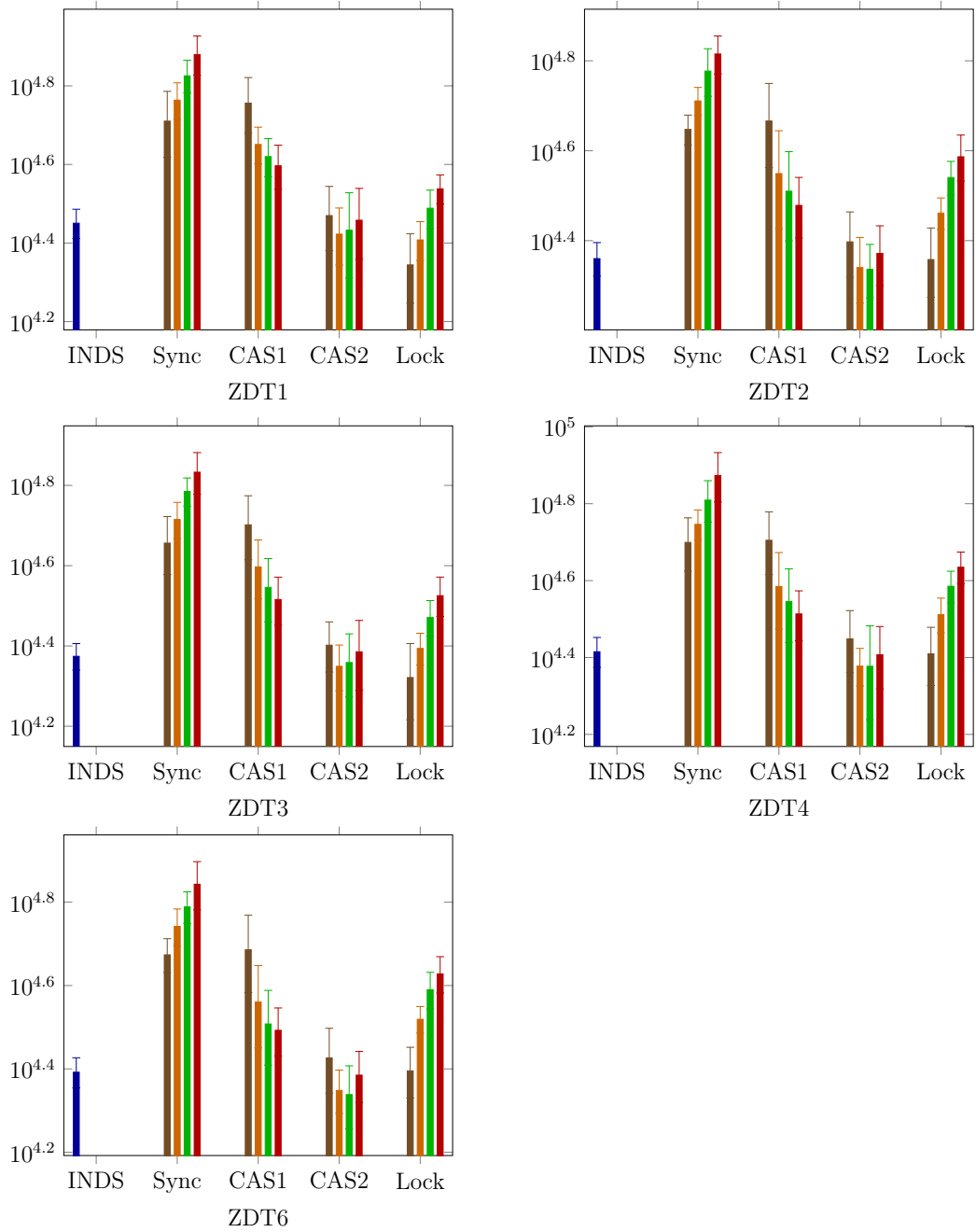

\begin{figure}[!t]
\begin{tikzpicture}[scale=0.8]
\includefivewide{\dtlzBdDtB}{\dtlzBdDtD}{\dtlzBdDtG}{\dtlzBdDtBC}{\dtlzBdDtCE}{DTLZ1}
\end{tikzpicture}
\begin{tikzpicture}[scale=0.8]
\includefivewide{\dtlzCdDtB}{\dtlzCdDtD}{\dtlzCdDtG}{\dtlzCdDtBC}{\dtlzCdDtCE}{DTLZ2}
\end{tikzpicture}
\begin{tikzpicture}[scale=0.8]
\includefivewide{\dtlzDdDtB}{\dtlzDdDtD}{\dtlzDdDtG}{\dtlzDdDtBC}{\dtlzDdDtCE}{DTLZ3}
\end{tikzpicture}
\begin{tikzpicture}[scale=0.8]
\includefivewide{\dtlzEdDtB}{\dtlzEdDtD}{\dtlzEdDtG}{\dtlzEdDtBC}{\dtlzEdDtCE}{DTLZ4}
\end{tikzpicture}
\begin{tikzpicture}[scale=0.8]
\includefivewide{\dtlzHdDtB}{\dtlzHdDtD}{\dtlzHdDtG}{\dtlzHdDtBC}{\dtlzHdDtCE}{DTLZ7}
\end{tikzpicture}
\caption{Experiments with DTLZ problems. For all the problems, $k = 3$.
For every asynchronous algorithm the columns represent the number of threads.
They are, left-to-right, 3, 6, 12, 24. Times are given in microseconds}
\label{dtlz3d}
\end{figure}
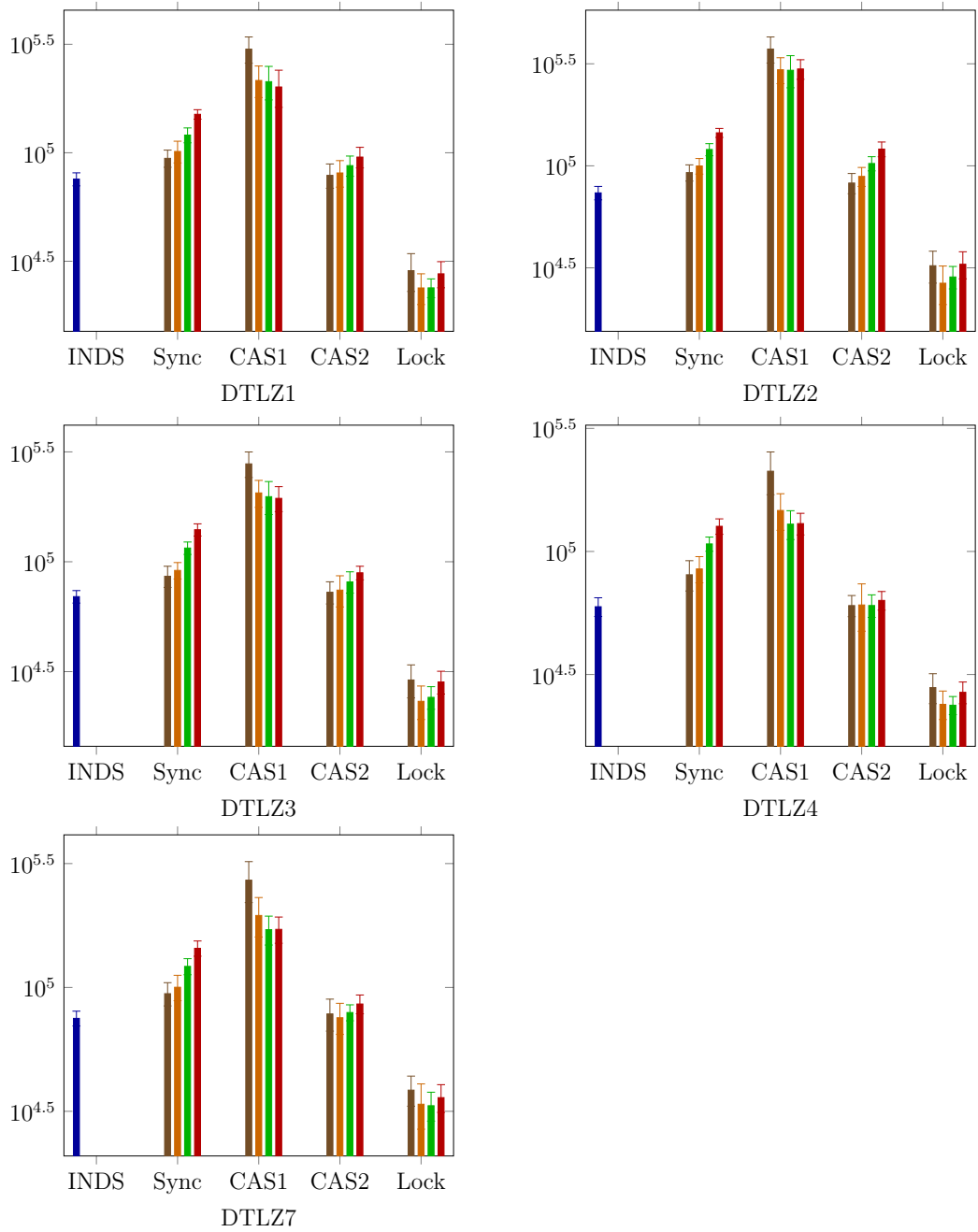

\begin{figure}[!t]
\begin{tikzpicture}[scale=0.8]
\includefivewide{\dtlzBdDtB}{\dtlzBdDtD}{\dtlzBdDtG}{\dtlzBdDtBC}{\dtlzBdDtCE}{DTLZ1, $k=3$}
\end{tikzpicture}
\begin{tikzpicture}[scale=0.8]
\includefivewide{\dtlzBdEtB}{\dtlzBdEtD}{\dtlzBdEtG}{\dtlzBdEtBC}{\dtlzBdEtCE}{DTLZ1, $k=4$}
\end{tikzpicture}
\begin{tikzpicture}[scale=0.8]
\includefivewide{\dtlzBdGtB}{\dtlzBdGtD}{\dtlzBdGtG}{\dtlzBdGtBC}{\dtlzBdGtCE}{DTLZ1, $k=6$}
\end{tikzpicture}
\begin{tikzpicture}[scale=0.8]
\includefivewide{\dtlzBdItB}{\dtlzBdItD}{\dtlzBdItG}{\dtlzBdItBC}{\dtlzBdItCE}{DTLZ1, $k=8$}
\end{tikzpicture}
\begin{tikzpicture}[scale=0.8]
\includefivewide{\dtlzBdBAtB}{\dtlzBdBAtD}{\dtlzBdBAtG}{\dtlzBdBAtBC}{\dtlzBdBAtCE}{DTLZ1, $k=10$}
\end{tikzpicture}
\caption{Experiments with DTLZ1 problem with varying $k$.
For every asynchronous algorithm the columns represent the number of threads.
They are, left-to-right, 3, 6, 12, 24. Times are given in microseconds}
\label{dtlz1}
\end{figure}
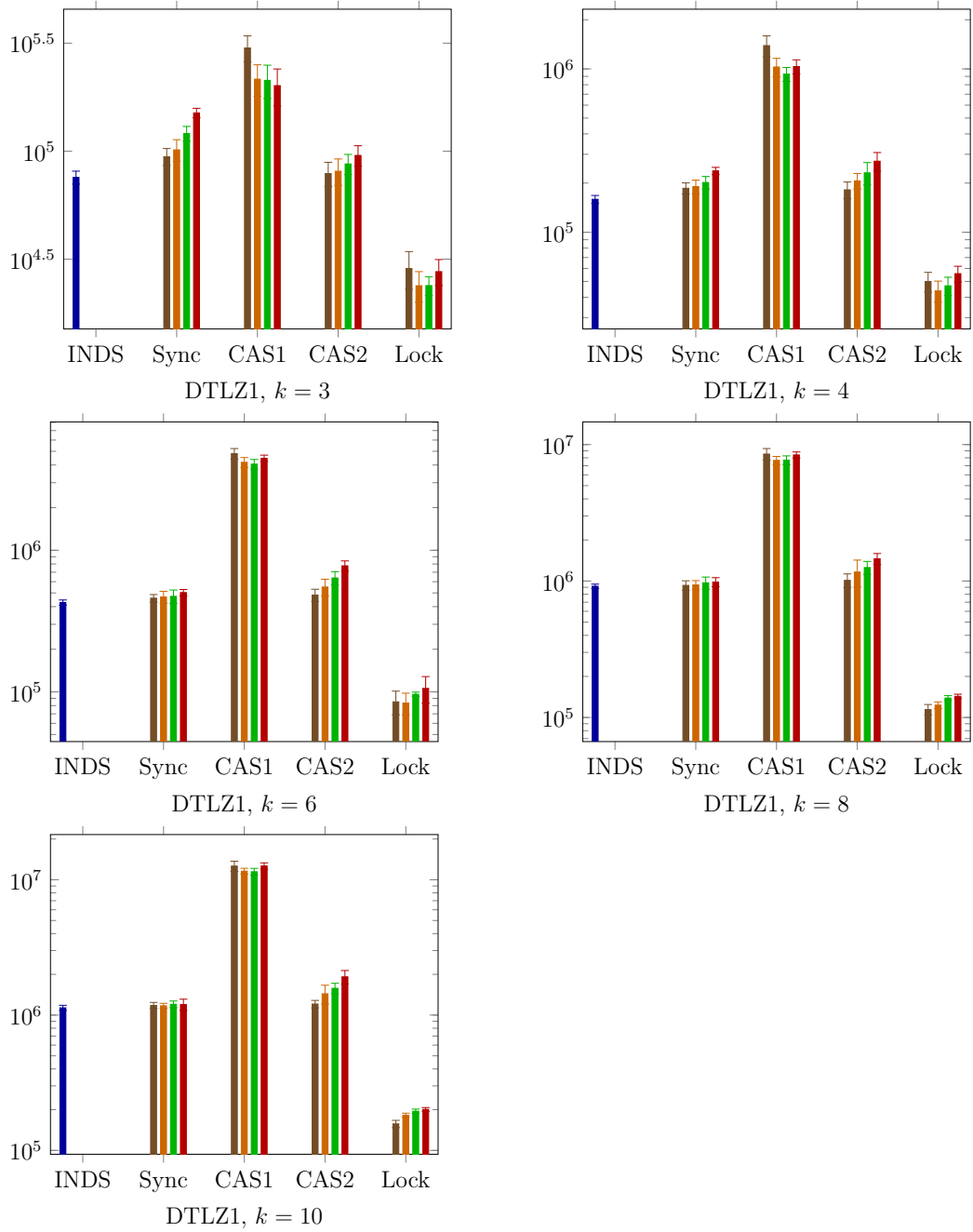

\begin{figure}[!t]
\begin{tikzpicture}[scale=0.8]
\includefivewide{\dtlzCdDtB}{\dtlzCdDtD}{\dtlzCdDtG}{\dtlzCdDtBC}{\dtlzCdDtCE}{DTLZ2, $k=3$}
\end{tikzpicture}
\begin{tikzpicture}[scale=0.8]
\includefivewide{\dtlzCdEtB}{\dtlzCdEtD}{\dtlzCdEtG}{\dtlzCdEtBC}{\dtlzCdEtCE}{DTLZ2, $k=4$}
\end{tikzpicture}
\begin{tikzpicture}[scale=0.8]
\includefivewide{\dtlzCdGtB}{\dtlzCdGtD}{\dtlzCdGtG}{\dtlzCdGtBC}{\dtlzCdGtCE}{DTLZ2, $k=6$}
\end{tikzpicture}
\begin{tikzpicture}[scale=0.8]
\includefivewide{\dtlzCdItB}{\dtlzCdItD}{\dtlzCdItG}{\dtlzCdItBC}{\dtlzCdItCE}{DTLZ2, $k=8$}
\end{tikzpicture}
\begin{tikzpicture}[scale=0.8]
\includefivewide{\dtlzCdBAtB}{\dtlzCdBAtD}{\dtlzCdBAtG}{\dtlzCdBAtBC}{\dtlzCdBAtCE}{DTLZ2, $k=10$}
\end{tikzpicture}
\caption{Experiments with DTLZ2 problem with varying $k$.
For every asynchronous algorithm the columns represent the number of threads.
They are, left-to-right, 3, 6, 12, 24. Times are given in microseconds.}
\label{dtlz2}
\end{figure}
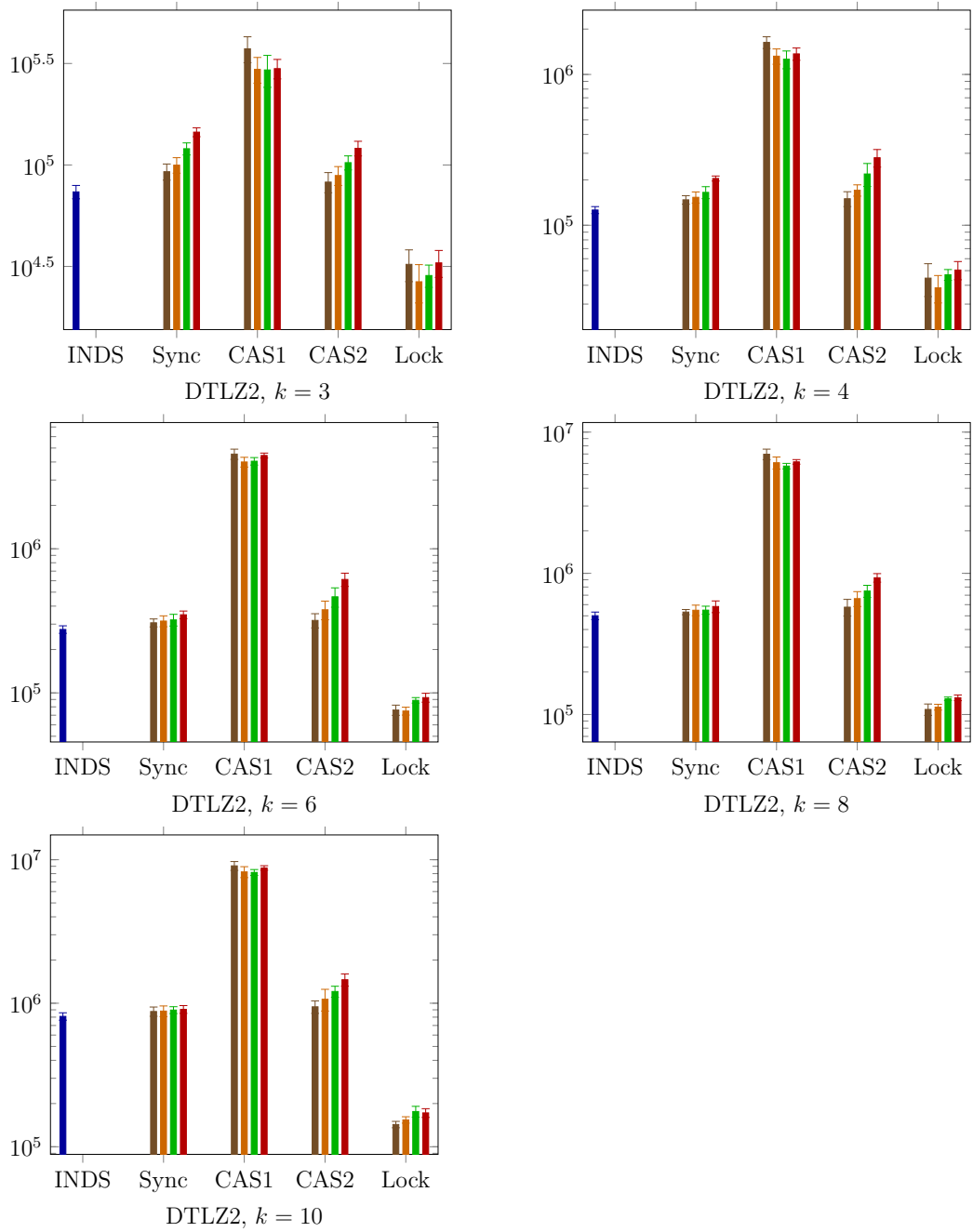

\end{document}

%% file: pic/tikz.tex
\newcommand{\convexpath}[2]{
[   
    create hullnodes/.code={
        \global\edef\namelist{#1}
        \foreach [count=\counter] \nodename in \namelist {
            \global\edef\numberofnodes{\counter}
            \node at (\nodename) [draw=none,name=hullnode\counter] {};
        }
        \node at (hullnode\numberofnodes) [name=hullnode0,draw=none] {};
        \pgfmathtruncatemacro\lastnumber{\numberofnodes+1}
        \node at (hullnode1) [name=hullnode\lastnumber,draw=none] {};
    },
    create hullnodes
]
($(hullnode1)!#2!-90:(hullnode0)$)
\foreach [
    evaluate=\currentnode as \previousnode using \currentnode-1,
    evaluate=\currentnode as \nextnode using \currentnode+1
    ] \currentnode in {1,...,\numberofnodes} {
-- ($(hullnode\currentnode)!#2!-90:(hullnode\previousnode)$)
  let \p1 = ($(hullnode\currentnode)!#2!-90:(hullnode\previousnode) - (hullnode\currentnode)$),
    \n1 = {atan2(\y1,\x1)},
    \p2 = ($(hullnode\currentnode)!#2!90:(hullnode\nextnode) - (hullnode\currentnode)$),
    \n2 = {atan2(\y2,\x2)},
    \n{delta} = {-Mod(\n1-\n2,360)}
  in 
    {arc [start angle=\n1, delta angle=\n{delta}, radius=#2]}
}
-- cycle
}

\newcommand{\ph}{\phantom{.}}

\newcommand{\TikZPictureIncrementalShot}{
    \draw[->] (0,0)--(11,0);
    \draw[->] (0,0)--(0,11);

    \node (aa) at (1, 4) {\ph};
    \node (ab) at (2, 2) {\ph};
    \node (ac) at (5, 1) {\ph};

    \node (ba) at (1, 7) {\ph};
    \node (bb) at (2, 6) {\ph};
    \node[moving1] (bc) at (4, 5) {\ph};
    \node[moving1] (bd) at (5, 4) {\ph};
    \node[moving1] (be) at (7, 3) {\ph};
    \node (bf) at (10, 2) {\ph};

    \node (ca) at (3, 10) {\ph};
    \node[moving2] (cb) at (4, 8) {\ph};
    \node[moving2] (cc) at (5, 6) {\ph};
    \node[moving2] (cd) at (6, 5) {\ph};
    \node[moving2] (ce) at (9, 3) {\ph};

    \node[moving3] (da) at (5, 9) {\ph};
    \node[moving3] (db) at (7, 8) {\ph};
    \node[moving3] (dc) at (8, 7) {\ph};

    \node[newnode] (new) at (3, 3) {\ph};
}

\newcommand{\TikZPictureLevel}{
    \tikzset{every node/.style={draw,circle,inner sep=2pt}}
    \begin{scope}[xshift=0cm,
                  newnode/.style={fill=gray},
                  moving1/.style={},
                  moving2/.style={},
                  moving3/.style={}]
        \TikZPictureIncrementalShot
        \draw (aa)--(ab)--(ac);
        \draw (da)--(db)--(dc);
        \draw (ba)--(bb)--(bc)--(bd)--(be)--(bf);
        \draw (ca)--(cb)--(cc)--(cd)--(ce);
        \draw[dashed] (new)-- ++(0,3) (new)-- ++(5,0);
    \end{scope}
    \draw[->] (11,5.5)--(12,5.5);
    \begin{scope}[xshift=13cm,
                  newnode/.style={fill=gray},
                  moving1/.style={fill=gray!70},
                  moving2/.style={},
                  moving3/.style={}]
        \TikZPictureIncrementalShot
        \draw (aa)--(ab)--(ac);
        \draw (ba)--(bb);
        \draw (ca)--(cb)--(cc)--(cd)--(ce);
        \draw (da)--(db)--(dc);
        \begin{pgfonlayer}{background}
            \draw (bb)--(bc) (be)--(bf);
            \draw[thick]   \convexpath{new,bc,bd,be}{6mm};
            \fill[gray!30] \convexpath{new,bc,bd,be}{6mm};
        \end{pgfonlayer}
    \end{scope}
    \draw[->] (24,5.5)--(25,5.5);
    \begin{scope}[xshift=26cm,
                  newnode/.style={},
                  moving1/.style={fill=gray},
                  moving2/.style={},
                  moving3/.style={}]
        \TikZPictureIncrementalShot
        \draw (aa)--(ab)--(ac);
        \draw (ba)--(bb)--(new)--(bf);
        \draw (bc)--(bd)--(be);
        \draw (ca)--(cb)--(cc)--(cd)--(ce);
        \draw (da)--(db)--(dc);
        \draw[dashed] (4,3)-- ++(0,6) (4,3)-- ++(6,0);
    \end{scope}
    \draw[->] (37,5.5)--(38,5.5);
    \begin{scope}[xshift=39cm,
                  newnode/.style={},
                  moving1/.style={fill=gray},
                  moving2/.style={fill=gray!70},
                  moving3/.style={}]
        \TikZPictureIncrementalShot
        \draw (aa)--(ab)--(ac);
        \draw (ba)--(bb)--(new)--(bf);
        \draw (da)--(db)--(dc);
        \begin{pgfonlayer}{background}
            \draw (ca)--(cb);
            \draw[thick]   \convexpath{cb,cc,cd,ce,be,bd,bc}{6mm};
            \fill[gray!30] \convexpath{cb,cc,cd,ce,be,bd,bc}{6mm};
        \end{pgfonlayer}
    \end{scope}
    \draw[->] (50,5.5)--(51,5.5);
    \begin{scope}[xshift=52cm,
                  newnode/.style={},
                  moving1/.style={},
                  moving2/.style={fill=gray},
                  moving3/.style={}]
        \TikZPictureIncrementalShot
        \draw (aa)--(ab)--(ac);
        \draw (ba)--(bb)--(new)--(bf);
        \draw (ca)--(bc)--(bd)--(be);
        \draw (cb)--(cc)--(cd)--(ce);
        \draw (da)--(db)--(dc);
        \draw[dashed] (4,3)-- ++(0,7) (4,3)-- ++(6,0);
    \end{scope}
}

%% file: pic/plots.tex
\pgfplotstableread{
  x y y-dev
  CAS2 218862.06699999998 37968.799026987836
  Sync 165233.891 15057.452253629512
  CAS1 1261406.334 170040.92167936606
  Lock 46987.996499999994 3875.21605685283
}{\dtlzCdEtBC}
\pgfplotstableread{
  x y y-dev
  INDS 59634.43374999999 5216.202524575006
}{\dtlzEdDtB}
\pgfplotstableread{
  x y y-dev
  INDS 916013.1145000001 33298.34122968368
}{\dtlzBdItB}
\pgfplotstableread{
  x y y-dev
  CAS1 1389557.53 204320.75493009246
  Lock 49928.5465 6884.693596854547
  CAS2 181896.00699999998 21229.65988048948
  Sync 185965.79450000002 14657.819552641944
}{\dtlzBdEtD}
\pgfplotstableread{
  x y y-dev
  Sync 531939.249 22454.61298523947
  CAS2 576101.7825 77927.56936350571
  Lock 108689.937 10013.759016115777
  CAS1 6977547.463500001 597281.6778371492
}{\dtlzCdItD}
\pgfplotstableread{
  x y y-dev
  Sync 582283.711 55253.524874304334
  CAS1 6158197.234 229916.85523658426
  CAS2 930485.6605 64711.325163667156
  Lock 131470.82749999998 6140.798452198297
}{\dtlzCdItCE}
\pgfplotstableread{
  x y y-dev
  Sync 55186.132666666665 5582.808638425556
  CAS1 36336.42633333334 8126.968327033806
  CAS2 22298.940666666662 2650.9735160035702
  Lock 33036.044 2418.1251266691447
}{\zdtGdCtG}
\pgfplotstableread{
  x y y-dev
  CAS2 1917852.5010000002 217542.97582078495
  Lock 200978.142 6501.0929910988425
  Sync 1196813.816 113665.73271002262
  CAS1 1.2647936416E7 689249.7724736138
}{\dtlzBdBAtCE}
\pgfplotstableread{
  x y y-dev
  Sync 51310.18400000001 9790.690717121903
  CAS2 29497.104000000003 5500.96951610532
  CAS1 57023.96 9217.507311779596
  Lock 22091.928333333333 4436.52474735699
}{\zdtBdCtD}
\pgfplotstableread{
  x y y-dev
  CAS2 120806.03833333333 10038.273577063937
  CAS1 298355.03066666663 32832.84134898204
  Sync 145179.30266666668 7350.132734881505
  Lock 32931.331999999995 4989.635331403521
}{\dtlzCdDtCE}
\pgfplotstableread{
  x y y-dev
  Lock 24763.219 2255.528627588442
  CAS1 39542.520333333334 6595.272065242241
  Sync 51879.43866666666 5365.804410279165
  CAS2 22344.90366666667 2911.898402165902
}{\zdtDdCtG}
\pgfplotstableread{
  x y y-dev
  INDS 75063.81862499999 5161.559655427708
}{\dtlzHdDtB}
\pgfplotstableread{
  x y y-dev
  CAS2 80729.36633333332 11278.150294174691
  CAS1 215399.31500000003 35729.631127542925
  Sync 101416.099 11643.767358276802
  Lock 23832.318 3853.67705783991
}{\dtlzBdDtG}
\pgfplotstableread{
  x y y-dev
  Sync 100202.06349999999 11781.07618204655
  CAS2 75493.543 10813.745521869283
  CAS1 195269.0145 35404.99017245719
  Lock 33791.278 7009.860947378379
}{\dtlzHdDtG}
\pgfplotstableread{
  x y y-dev
  CAS1 194386.47199999998 25318.12320464681
  CAS2 89085.509 6326.059695314217
  Lock 28377.74 3372.888291384156
  Sync 139870.3753333333 8941.99269382103
}{\dtlzDdDtCE}
\pgfplotstableread{
  x y y-dev
  CAS1 7662738.9785 529160.8601567502
  Sync 936442.372 70984.60164513116
  Lock 123222.185 6929.468120503802
  CAS2 1164370.059 259971.79703892308
}{\dtlzBdItG}
\pgfplotstableread{
  x y y-dev
  CAS2 26455.142666666667 4380.780312574538
  Lock 25575.971666666665 2892.6054743158456
  Sync 58033.83599999999 6195.931750515387
  CAS1 44749.54633333333 4787.744553492804
}{\zdtBdCtG}
\pgfplotstableread{
  x y y-dev
  Sync 74728.39199999999 10964.067267007395
  Lock 43129.71233333334 4096.138360191015
  CAS2 25539.312333333335 4698.971669247574
  CAS1 32608.20533333333 4833.433456393877
}{\zdtEdCtCE}
\pgfplotstableread{
  x y y-dev
  Lock 27681.747333333333 3809.3065541842825
  CAS1 201046.13066666666 39126.91621653204
  CAS2 95550.57133333334 10411.750485463366
  Sync 150420.482 7547.1675498319455
}{\dtlzBdDtCE}
\pgfplotstableread{
  x y y-dev
  INDS 126212.17049999998 6752.189467921349
}{\dtlzCdEtB}
\pgfplotstableread{
  x y y-dev
  Lock 25670.297666666665 4426.7698992406795
  CAS1 50678.454333333335 9426.833683890862
  Sync 50044.920666666665 7935.345400834462
  CAS2 28089.232333333333 5171.329842310261
}{\zdtEdCtD}
\pgfplotstableread{
  x y y-dev
  INDS 427607.6547499999 18206.329577118777
}{\dtlzBdGtB}
\pgfplotstableread{
  x y y-dev
  CAS2 23850.773333333334 2673.542687961799
  Lock 32482.096999999998 3365.6333076314672
  CAS1 38429.308 8656.517879143033
  Sync 55756.36033333334 5025.250426062499
}{\zdtEdCtG}
\pgfplotstableread{
  x y y-dev
  Sync 66919.18 6382.641982696716
  CAS2 27095.413666666664 6626.134638910028
  CAS1 41672.695999999996 4639.283535514078
  Lock 30801.058333333334 3453.5982184885083
}{\zdtBdCtBC}
\pgfplotstableread{
  x y y-dev
  Lock 172195.626 12020.624169632832
  CAS1 8767149.844999999 316872.12974532647
  CAS2 1457979.271 142588.0135120188
  Sync 906505.1059999999 56473.54140535757
}{\dtlzCdBAtCE}
\pgfplotstableread{
  x y y-dev
  INDS 69409.03391666668 4537.9926144510655
}{\dtlzDdDtB}
\pgfplotstableread{
  x y y-dev
  Sync 320976.4525 29991.13395432344
  CAS2 464510.963 71138.67919345749
  Lock 88847.77100000001 3944.0900534478164
  CAS1 4052478.4705 242211.44199945146
}{\dtlzCdGtBC}
\pgfplotstableread{
  x y y-dev
  Sync 59856.583999999995 7268.738496645733
  CAS1 32355.630333333334 7273.128228260931
  Lock 34696.88333333333 3001.367118480399
  CAS2 21692.296 2945.032845875011
}{\zdtCdCtBC}
\pgfplotstableread{
  x y y-dev
  Sync 1179622.8560000001 60891.72819132747
  CAS1 1.26211532835E7 1087654.3451888035
  Lock 157118.94650000002 9705.803814907167
  CAS2 1207394.638 76477.49998767485
}{\dtlzBdBAtD}
\pgfplotstableread{
  x y y-dev
  CAS2 272375.63049999997 35316.71178694309
  Sync 237567.238 12138.640508390325
  Lock 55795.047 6126.020136658465
  CAS1 1034208.091 102944.8214099293
}{\dtlzBdEtCE}
\pgfplotstableread{
  x y y-dev
  Lock 20954.032333333336 4509.034658713769
  CAS1 50316.48166666667 9141.481671100752
  Sync 45300.981999999996 7459.506696315581
  CAS2 25230.95 3604.5009104856204
}{\zdtDdCtD}
\pgfplotstableread{
  x y y-dev
  CAS2 170703.1485 14702.415186786167
  CAS1 1322456.3125 153906.03294387241
  Lock 38476.239499999996 7943.561195788511
  Sync 153291.592 12998.730624741016
}{\dtlzCdEtG}
\pgfplotstableread{
  x y y-dev
  CAS1 929437.7765 91820.56287084919
  Lock 47004.5975 6038.5568329634025
  Sync 201748.46649999998 17727.397491870288
  CAS2 231040.64649999997 36098.287021850614
}{\dtlzBdEtBC}
\pgfplotstableread{
  x y y-dev
  CAS2 749309.9665000001 72813.60787561299
  Sync 549018.8895 36140.66248904929
  CAS1 5748718.4515 256263.69981490227
  Lock 129356.53 3611.2023339394873
}{\dtlzCdItBC}
\pgfplotstableread{
  x y y-dev
  Sync 147493.71899999998 9581.5885347502
  CAS1 1631991.9695000001 146076.88382206162
  CAS2 149955.15899999999 16963.86472535868
  Lock 44628.1535 10963.363054547382
}{\dtlzCdEtD}
\pgfplotstableread{
  x y y-dev
  Sync 61459.51066666667 5331.313506025321
  CAS2 21792.67466666667 3752.5933069862504
  CAS1 32195.649333333335 6555.873186257622
  Lock 38901.543333333335 3918.44996418452
}{\zdtGdCtBC}
\pgfplotstableread{
  x y y-dev
  INDS 275525.457875 16580.99723864798
}{\dtlzCdGtB}
\pgfplotstableread{
  x y y-dev
  INDS 809896.241125 47973.72067411074
}{\dtlzCdBAtB}
\pgfplotstableread{
  x y y-dev
  CAS2 87259.11866666666 9319.551445027062
  Lock 23854.258 2370.4440241126413
  CAS1 212668.982 37409.68772823687
  Sync 120663.71433333332 9508.756484264999
}{\dtlzBdDtBC}
\pgfplotstableread{
  x y y-dev
  CAS2 21884.447666666663 3647.57555856791
  CAS1 35407.051 8695.509573737623
  Sync 51371.343 3683.082169981142
  Lock 28929.476333333336 2327.0104313768484
}{\zdtCdCtG}
\pgfplotstableread{
  x y y-dev
  Sync 44399.416333333334 3378.9383113112617
  CAS2 24962.888666666666 4131.182620119531
  CAS1 46348.89366666666 9840.355538555234
  Lock 22799.165999999997 4005.0987558393613
}{\zdtCdCtD}
\pgfplotstableread{
  x y y-dev
  Sync 143894.446 10173.344445565037
  Lock 35919.9215 4604.367446082903
  CAS2 85772.2185 7459.350678201957
  CAS1 171606.446 20832.71070340359
}{\dtlzHdDtCE}
\pgfplotstableread{
  x y y-dev
  Sync 47163.124 4401.336367462728
  CAS1 48498.032999999996 10200.064796045088
  CAS2 26699.856 4729.514393767786
  Lock 24847.838333333333 3453.9544059193468
}{\zdtGdCtD}
\pgfplotstableread{
  x y y-dev
  CAS2 1011744.2904999999 116483.25680341315
  CAS1 8524782.2565 829959.2116003822
  Lock 114446.99799999999 10041.593089538756
  Sync 929858.4955 74414.87169476466
}{\dtlzBdItD}
\pgfplotstableread{
  x y y-dev
  Sync 94293.24466666667 8565.668422264022
  CAS1 300314.974 41516.409493118575
  Lock 28622.183666666668 5668.453283310624
  CAS2 78707.03833333333 10089.890134103229
}{\dtlzBdDtD}
\pgfplotstableread{
  x y y-dev
  CAS1 129107.50066666666 17329.7108491353
  Sync 107308.81733333335 7116.779330967836
  Lock 23725.830333333335 2009.2633696208998
  CAS2 60299.34566666667 6334.941187830582
}{\dtlzEdDtBC}
\pgfplotstableread{
  x y y-dev
  Sync 80406.49433333334 11320.718327542852
  CAS1 211893.96766666663 41872.57987521004
  CAS2 60274.307 5945.2248455594745
  Lock 27970.71533333333 3893.5795744006996
}{\dtlzEdDtD}
\pgfplotstableread{
  x y y-dev
  INDS 159138.50125 8857.961210375783
}{\dtlzBdEtB}
\pgfplotstableread{
  x y y-dev
  Lock 42445.494999999995 4240.559455204152
  CAS2 24262.912 3394.6697694424806
  Sync 69657.953 9271.321828236756
  CAS1 31085.463666666667 4087.4193352524603
}{\zdtGdCtCE}
\pgfplotstableread{
  x y y-dev
  Sync 984110.066 74913.32648100563
  CAS2 1457467.803 133807.99039367284
  CAS1 8402329.907 433676.8675265368
  Lock 142971.3375 5183.921926402104
}{\dtlzBdItCE}
\pgfplotstableread{
  x y y-dev
  CAS1 293983.07033333334 52697.12995995982
  CAS2 102670.35599999999 8194.196434910626
  Sync 120254.74766666668 8199.164868013937
  Lock 28536.515 3588.3145301839136
}{\dtlzCdDtBC}
\pgfplotstableread{
  x y y-dev
  CAS1 4180359.5115 337337.83687533694
  Lock 83350.468 14592.642504421534
  Sync 467925.5105 45453.56784449284
  CAS2 550114.9155 74343.21273920317
}{\dtlzBdGtG}
\pgfplotstableread{
  x y y-dev
  Sync 472693.26599999995 51505.01812725383
  Lock 96238.5815 3438.519781953275
  CAS2 637363.7185 68580.85626451945
  CAS1 4057441.8405 319037.5514994899
}{\dtlzBdGtBC}
\pgfplotstableread{
  x y y-dev
  CAS1 1369612.7375 127750.83240454945
  Lock 50483.788 6987.509210440191
  Sync 204314.3045 7851.027646792999
  CAS2 280432.948 37440.29749484812
}{\dtlzCdEtCE}
\pgfplotstableread{
  x y y-dev
  INDS 23666.883416666667 1796.7147997046118
}{\zdtDdCtB}
\pgfplotstableread{
  x y y-dev
  CAS2 483046.5325 47781.440146646084
  Sync 459004.3855 27881.263928912595
  Lock 85082.52299999999 16163.445851461562
  CAS1 4815741.8155000005 411831.11484680546
}{\dtlzBdGtD}
\pgfplotstableread{
  x y y-dev
  INDS 73707.18116666666 5583.824513364205
}{\dtlzCdDtB}
\pgfplotstableread{
  x y y-dev
  Sync 84930.00666666667 10354.453804843048
  CAS2 60621.96633333334 13205.525782097142
  Lock 23921.104333333333 3169.923000499644
  CAS1 146775.67333333334 24890.53389141193
}{\dtlzEdDtG}
\pgfplotstableread{
  x y y-dev
  INDS 28207.328916666665 2410.16922011832
}{\zdtBdCtB}
\pgfplotstableread{
  x y y-dev
  Lock 29563.911666666667 3020.2165078072353
  Sync 60929.272 4910.971723992038
  CAS2 22838.299 4084.8580377386024
  CAS1 35146.57833333333 6306.192806990575
}{\zdtDdCtBC}
\pgfplotstableread{
  x y y-dev
  Lock 26593.792666666664 5693.586811038773
  CAS1 295762.869 42947.11367229149
  CAS2 88731.393 9456.90340668759
  Sync 99848.84266666666 8771.300341365166
}{\dtlzCdDtG}
\pgfplotstableread{
  x y y-dev
  CAS2 662033.8495 80160.07796562847
  Sync 548150.3859999999 47081.35777539514
  Lock 112862.853 5338.096261187597
  CAS1 6071588.254000001 595871.4372460903
}{\dtlzCdItG}
\pgfplotstableread{
  x y y-dev
  Sync 306514.361 19469.287780750994
  CAS1 4541830.83 370068.65157333045
  CAS2 318469.033 35958.45345921477
  Lock 76048.33499999999 6196.29002012192
}{\dtlzCdGtD}
\pgfplotstableread{
  x y y-dev
  Lock 43750.623999999996 6474.128891501388
  CAS2 206293.994 22791.295943164903
  CAS1 1027518.3744999999 133598.7892690795
  Sync 190315.2355 18254.544648451367
}{\dtlzBdEtG}
\pgfplotstableread{
  x y y-dev
  Lock 33324.898 4449.325161033074
  CAS1 171178.3895 22861.985985194387
  CAS2 79239.8725 5865.021854462692
  Sync 121631.7595 9045.019702836391
}{\dtlzHdDtBC}
\pgfplotstableread{
  x y y-dev
  INDS 1131479.3401249999 50340.63288066082
}{\dtlzBdBAtB}
\pgfplotstableread{
  x y y-dev
  Lock 105821.4375 22456.847843709646
  Sync 502755.5735 25854.03067067899
  CAS1 4461016.154999999 236561.59784252485
  CAS2 776884.2185 64298.51363753275
}{\dtlzBdGtCE}
\pgfplotstableread{
  x y y-dev
  CAS1 129819.20833333333 13083.266018035736
  Lock 26769.291999999998 2711.1679581034814
  CAS2 63262.120333333325 5497.7848423466585
  Sync 126553.76033333334 9111.284255047583
}{\dtlzEdDtCE}
\pgfplotstableread{
  x y y-dev
  CAS1 9069507.298999999 642851.1333158785
  Lock 142976.668 7272.17622961614
  CAS2 944417.996 94245.48280640662
  Sync 876200.4945 65012.81073880066
}{\dtlzCdBAtD}
\pgfplotstableread{
  x y y-dev
  Sync 75825.65566666667 8700.030020854449
  CAS2 28705.091666666664 5894.649182898588
  Lock 34514.903 2930.2809106383183
  CAS1 39496.937666666665 5069.6988912768775
}{\zdtBdCtCE}
\pgfplotstableread{
  x y y-dev
  Sync 1202874.0255 68205.30135227507
  Lock 194922.448 7202.258818295167
  CAS1 1.1464476928E7 674253.6723693121
  CAS2 1577824.095 141263.22064112
}{\dtlzBdBAtBC}
\pgfplotstableread{
  x y y-dev
  Lock 74872.9855 4716.934815646545
  CAS2 377515.8895 55866.012302899966
  Sync 315126.44350000005 27429.5912631268
  CAS1 4002258.0154999997 311507.5019454171
}{\dtlzCdGtG}
\pgfplotstableread{
  x y y-dev
  INDS 25998.563750000005 2323.644997304343
}{\zdtEdCtB}
\pgfplotstableread{
  x y y-dev
  CAS1 7700094.2905 569209.4253531744
  Sync 971010.1835 98965.94049810042
  CAS2 1256271.6570000001 136868.42947707587
  Lock 138860.072 5751.217423683529
}{\dtlzBdItBC}
\pgfplotstableread{
  x y y-dev
  CAS1 271081.145 51029.49699773284
  Lock 38530.104999999996 5358.130443849468
  Sync 94403.8675 10154.424999406909
  CAS2 78198.3185 11532.70183351087
}{\dtlzHdDtD}
\pgfplotstableread{
  x y y-dev
  Sync 348667.003 21306.66406847078
  Lock 92779.70850000001 6606.116803653187
  CAS2 612776.6585 64596.459733288604
  CAS1 4441239.3255 173765.2824081257
}{\dtlzCdGtCE}
\pgfplotstableread{
  x y y-dev
  CAS2 1066276.412 184027.88014264824
  CAS1 8226714.4655 718304.5417265439
  Lock 153929.858 7900.9713799230085
  Sync 882075.3095 75550.0676390183
}{\dtlzCdBAtG}
\pgfplotstableread{
  x y y-dev
  CAS1 197867.96333333335 33908.45002692373
  Sync 115494.72000000002 7509.820625693732
  Lock 24163.744666666666 2848.813223972397
  CAS2 81001.35366666668 9004.708020208465
}{\dtlzDdDtBC}
\pgfplotstableread{
  x y y-dev
  Lock 38489.583333333336 3623.3767344870453
  CAS1 35106.20266666667 7605.470360487224
  Sync 64467.513 7956.698432096317
  CAS2 23829.506000000005 6531.029126885109
}{\zdtEdCtBC}
\pgfplotstableread{
  x y y-dev
  CAS2 24283.97 4797.8671600062735
  Lock 33518.23533333333 3774.738427352285
  Sync 68048.82433333334 8114.753251786813
  CAS1 32787.305 4489.898140771198
}{\zdtDdCtCE}
\pgfplotstableread{
  x y y-dev
  Sync 1171441.696 50219.21943884686
  Lock 182044.195 6449.286451893186
  CAS2 1436822.7289999998 225606.41762089412
  CAS1 1.1585631276E7 568655.2335308081
}{\dtlzBdBAtG}
\pgfplotstableread{
  x y y-dev
  CAS2 1207744.4255 105857.66532099192
  CAS1 8159955.091 385029.4273331037
  Sync 896517.6545 49475.39697483146
  Lock 176060.79200000002 15246.833059265684
}{\dtlzCdBAtBC}
\pgfplotstableread{
  x y y-dev
  INDS 75604.49983333334 5212.670934498815
}{\dtlzBdDtB}
\pgfplotstableread{
  x y y-dev
  Sync 92646.05533333332 8464.399287922899
  CAS1 373599.93333333335 54586.4616766877
  CAS2 82351.46266666667 9375.328267549125
  Lock 32395.464333333333 5775.857422502769
}{\dtlzCdDtD}
\pgfplotstableread{
  x y y-dev
  Sync 91344.586 7831.968381388211
  CAS2 74309.469 12034.645878120415
  CAS1 205728.96433333334 28873.18683140609
  Lock 23175.27733333333 4027.1085034062594
}{\dtlzDdDtG}
\pgfplotstableread{
  x y y-dev
  INDS 22908.508583333332 1960.2674031149213
}{\zdtCdCtB}
\pgfplotstableread{
  x y y-dev
  CAS2 23524.721 3569.5507615051506
  Sync 65327.13833333333 6322.842028691686
  CAS1 30089.751 4636.059224222264
  Lock 38595.75433333334 4532.977452003852
}{\zdtCdCtCE}
\pgfplotstableread{
  x y y-dev
  Sync 85930.29299999999 9492.388490332996
  CAS1 278965.3796666667 37129.30272330994
  CAS2 72672.72733333333 8391.708101772963
  Lock 28978.13666666667 4961.535134093446
}{\dtlzDdDtD}
\pgfplotstableread{
  x y y-dev
  INDS 24657.897583333335 2050.121266618067
}{\zdtGdCtB}
\pgfplotstableread{
  x y y-dev
  INDS 500895.40612500004 30811.635114343666
}{\dtlzCdItB}

%% file: async-nds.bbl
\begin{thebibliography}{10}

\bibitem{hype-algorithm}
J.~Bader and E.~Zitzler.
\newblock {HypE}: An algorithm for fast hypervolume-based many-objective
  optimization.
\newblock {\em Evolutionary Computation}, 19(1):45--76, 2011.

\bibitem{buzdalov-nds-2014}
M.~Buzdalov and A.~Shalyto.
\newblock A provably asymptotically fast version of the generalized {J}ensen
  algorithm for non-dominated sorting.
\newblock In {\em Parallel Problem Solving from Nature -- {PPSN} {XIII}},
  number 8672 in Lecture Notes in Computer Science, pages 528--537. Springer,
  2014.

\bibitem{sync-async-moea}
A.~W. Churchill, P.~Husbands, and A.~Philippides.
\newblock Tool sequence optimization using synchronous and asynchronous
  parallel multi-objective evolutionary algorithms with heterogeneous
  evaluations.
\newblock In {\em Proceedings of IEEE Congress on Evolutionary Computation},
  pages 2924--2931, 2013.

\bibitem{pesa-ii}
D.~W. Corne, N.~R. Jerram, J.~D. Knowles, and M.~J. Oates.
\newblock {PESA}-{II}: Region-based selection in evolutionary multiobjective
  optimization.
\newblock In {\em Proceedings of Genetic and Evolutionary Computation
  Conference}, pages 283--290. Morgan Kaufmann Publishers, 2001.

\bibitem{nsga-iii}
K.~Deb and H.~Jain.
\newblock An evolutionary many-objective optimization algorithm using
  reference-point-based nondominated sorting approach, part {I}: Solving
  problems with box constraints.
\newblock {\em IEEE Transactions on Evolutionary Computation}, 18(4):577--601,
  2013.

\bibitem{nsga-ii}
K.~Deb, A.~Pratap, S.~Agarwal, and T.~Meyarivan.
\newblock A fast and elitist multi-objective genetic algorithm: {NSGA}-{II}.
\newblock {\em IEEE Transactions on Evolutionary Computation}, 6(2):182--197,
  2002.

\bibitem{dtlz}
K.~Deb, L.~Thiele, M.~Laumanns, and E.~Zitzler.
\newblock Scalable test problems for evolutionary multiobjective optimization.
\newblock In {\em Evolutionary Multiobjective Optimization. Theoretical
  Advances and Applications}, pages 105--145. Springer, 2005.

\bibitem{nebro-durillo-master-slave-nsga-ii}
J.~J. Durillo, A.~J. Nebro, F.~Luna, and E.~Alba.
\newblock A study of master-slave approaches to parallelize {NSGA-II}.
\newblock In {\em Proceedings of IEEE International Symposium on Parallel and
  Distributed Processing}, pages 1--8, 2008.

\bibitem{jensen}
M.~T. Jensen.
\newblock Reducing the run-time complexity of multiobjective {EA}s: The
  {NSGA}-{II} and other algorithms.
\newblock {\em IEEE Transactions on Evolutionary Computation}, 7(5):503--515,
  2003.

\bibitem{wait-free-queues}
A.~Kogan and E.~Petrank.
\newblock Wait-free queues with multiple enqueuers and dequeuers.
\newblock In {\em Proceedings of the ACM SIGPLAN Symposium on Principles and
  Practice of Parallel Programming}, pages 223--234, 2011.

\bibitem{deb-enlu-14}
K.~Li, K.~Deb, Q.~Zhang, and S.~Kwong.
\newblock Efficient non-domination level update approach for steady-state
  evolutionary multiobjective optimization.
\newblock Technical Report COIN 2014014, Michigan State University, 2014.

\bibitem{deb-enlu}
K.~Li, K.~Deb, Q.~Zhang, and Q.~Zhang.
\newblock Efficient nondomination level update method for steady-state
  evolutionary multiobjective optimization.
\newblock {\em IEEE Transactions on Cybernetics}, 47(9):2838--2849, 2017.

\bibitem{mishra-non-dominated-level-update}
S.~Mishra, S.~Mondal, and S.~Saha.
\newblock Improved solution to the non-domination level update problem.
\newblock {\em Applied Soft Computing}, 60:336--362, 2017.

\bibitem{nsga-ii-steady-state}
A.~J. Nebro and J.~J. Durillo.
\newblock On the effect of applying a steady-state selection scheme in the
  multi-objective genetic algorithm {NSGA}-{II}.
\newblock In {\em Nature-Inspired Algorithms for Optimisation}, number 193 in
  Studies in Computational Intelligence, pages 435--456. Springer Berlin
  Heidelberg, 2009.

\bibitem{async-master-slave-moea-heterocosts}
M.~Yagoubi and M.~Schoenauer.
\newblock Asynchronous master/slave {MOEA}s and heterogeneous evaluation costs.
\newblock In {\em Proceedings of Genetic and Evolutionary Computation
  Conference}, pages 1007--1014, 2012.

\bibitem{async-moea-heterocosts}
M.~Yagoubi, L.~Thobois, and M.~Schoenauer.
\newblock Asynchronous evolutionary multi-objective algorithms with
  heterogeneous evaluation costs.
\newblock In {\em Proceedings of IEEE Congress on Evolutionary Computation},
  pages 21--28, 2011.

\bibitem{incremental-nds-cec15}
I.~Yakupov and M.~Buzdalov.
\newblock Incremental non-dominated sorting with {$O(N)$} insertion for the
  two-dimensional case.
\newblock In {\em Proceedings of IEEE Congress on Evolutionary Computation},
  pages 1853--1860, 2015.

\bibitem{yakupovB-gecco17-inds}
I.~Yakupov and M.~Buzdalov.
\newblock Improved incremental non-dominated sorting for steady-state
  evolutionary multiobjective optimization.
\newblock In {\em Proceedings of Genetic and Evolutionary Computation
  Conference}, pages 649--656, 2017.

\bibitem{moea-d}
Q.~Zhang and H.~Li.
\newblock {MOEA/D}: A multiobjective evolutionary algorithm based on
  decomposition.
\newblock {\em IEEE Transactions on Evolutionary Computation}, 11(6):712--731,
  2007.

\bibitem{zdt}
E.~Zitzler, K.~Deb, and L.~Thiele.
\newblock Comparison of multiobjective evolutionary algorithms: Empirical
  results.
\newblock {\em Evolutionary Computation}, 8(2):173--195, 2000.

\bibitem{ibea}
E.~Zitzler and S.~K{\"u}nzli.
\newblock Indicator-based selection in multiobjective search.
\newblock In {\em Parallel Problem Solving from Nature -- {PPSN} {VIII}},
  number 3242 in Lecture Notes in Computer Science, pages 832--842. 2004.

\bibitem{spea2}
E.~Zitzler, M.~Laumanns, and L.~Thiele.
\newblock {SPEA2}: Improving the strength pareto evolutionary algorithm for
  multiobjective optimization.
\newblock In {\em Proceedings of the EUROGEN'2001 Conference}, pages 95--100,
  2001.

\bibitem{hypervolume}
E.~Zitzler and L.~Thiele.
\newblock Multiobjective evolutionary algorithms: A comparative case study and
  the {S}trength {P}areto approach.
\newblock {\em IEEE Transactions on Evolutionary Computation}, 3(4):257--271,
  1999.

\end{thebibliography}
